\documentclass[10pt,aps,prd,twocolumn,amsmath,amssymb,superscriptaddress,nofootinbib]{revtex4-2}

\usepackage{graphicx}
\usepackage{subcaption}
\usepackage{epsfig}
\usepackage{ragged2e}
\usepackage{bm}
\usepackage[normalem]{ulem}
\usepackage{color}
\usepackage{multirow}
\usepackage{slashed}
\usepackage{verbatim}
\usepackage{url}
\usepackage{braket}
\usepackage{amssymb}
\usepackage{adjustbox}

\usepackage{tikz}
\usetikzlibrary{arrows,shapes}
\usetikzlibrary{trees}
\usetikzlibrary{matrix,arrows} 				
\usetikzlibrary{positioning}				
\usetikzlibrary{calc,through}				
\usetikzlibrary{decorations.pathreplacing}  
\usepackage{pgffor}							

\usetikzlibrary{decorations.pathmorphing}	
\usetikzlibrary{decorations.markings}
\tikzset{
	vector/.style={decorate, decoration={snake}, draw},
	provector/.style={decorate, decoration={snake,amplitude=2.5pt}, draw},
	antivector/.style={decorate, decoration={snake,amplitude=-2.5pt}, draw},
	fermion/.style={draw=black, postaction={decorate},
		decoration={markings,mark=at position .55 with {\arrow[draw=black]{>}}}},
	fermiona/.style={draw=red},
	fermionbar/.style={draw=black, postaction={decorate},
		decoration={markings,mark=at position .55 with {\arrow[draw=black]{<}}}},
	fermionnoarrow/.style={draw=black},
	gluon/.style={decorate, draw=black,
		decoration={coil,amplitude=4pt, segment length=5pt}},
	scalar/.style={dashed,draw=black, postaction={decorate},
		decoration={markings,mark=at position .55 with {\arrow[draw=black]{>}}}},
	scalarbar/.style={dashed,draw=black, postaction={decorate},
		decoration={markings,mark=at position .55 with {\arrow[draw=black]{<}}}},
	scalarnoarrow/.style={dashed,draw=black},
	electron/.style={draw=black, postaction={decorate},
		decoration={markings,mark=at position .55 with {\arrow[draw=black]{>}}}},
	bigvector/.style={decorate, decoration={snake,amplitude=4pt}, draw},
}

\def\ben{\begin{equation}}
	\def\een{\end{equation}}

\def\be{\begin{equation}}
	\def\ee{\end{equation}}
\def\beq{\begin{equation}}
	\def\eeq{\end{equation}}
\def\ba{\begin{array}}
	\def\ea{\end{array}}

\def\dalemb#1#2{{\vbox{\hrule height .#2pt
			\hbox{\vrule width.#2pt height#1pt \kern#1pt
				\vrule width.#2pt}
			\hrule height.#2pt}}}

\newcommand{\bea}{\begin{eqnarray}}
	\newcommand{\eea}{\end{eqnarray}}

\newcommand{\Tr}{{\rm Tr} }

\DeclareDocumentCommand{\nint}{ O{} O{} m }{\ensuremath{ \int_{\mbox{\scriptsize $#1$}}^{\mbox{\scriptsize$#2$}}\!\!\! \mbox{\small $\,\mathrm{d}#3$\! }}}

\usepackage{hyperref}
\begin{document}

\title{Symmetry restoration in a fast scrambling system}

\author{Sizheng Cao}
\affiliation{Department of Physics, College of Sciences, Shanghai University, 99 Shangda Road, 200444 Shanghai, China}

\author{Xian-Hui Ge}
\email{gexh@shu.edu.cn}
\affiliation{Department of Physics, College of Sciences, Shanghai University, 99 Shangda Road,
200444 Shanghai, China}


\begin{abstract}
Entanglement asymmetry---used here as a direct probe of symmetry restoration---provides a sharp diagnostic of post-quench dynamics. We test this idea in the complex Sachdev-Ye-Kitaev model with a conserved U(1) charge. Using exact diagonalization, we track the joint evolution of entanglement entropy and entanglement asymmetry after quenches from charge-asymmetric product states. We find rapid volume-law entanglement growth consistent with the subsystem eigenstate thermalization hypothesis, accompanied by a concurrent decay of entanglement asymmetry to a late-time plateau set by finite-size effects: small subsystems display near-complete restoration, while residual cross-sector weight yields a finite plateau. Notably, we uncover a quantum Mpemba effect: states prepared further from symmetry relax faster and approach lower residual asymmetry; disorder in the couplings renders this behavior more robust and monotonic across parameters. We further derive a Pinsker-type lower bound that ties the decay of asymmetry to differences in subsystem purity, identifying dephasing between U(1) charge sectors as the operative mechanism. These results establish entanglement asymmetry as a sensitive probe of symmetry restoration and thermalization, clarifying finite-size limits in fast-scrambling, closed quantum systems.
\end{abstract}

\maketitle
\thanks{Corresponding author: gexh@shu.edu.cn}

\section{Introduction}

The thermalization of quantum systems \cite{D'Alessio03052016,Deutsch_2018,Gogolin_2016,RevModPhys.76.1267} has long been a core topic in the study of non-equilibrium dynamics. Meanwhile, there have been many studies on the connection between symmetry and thermalization \cite{Ares2023,SciPostPhys.15.3.089,Liushuo2024,Chalas_2024,PhysRevB.94.224206,PhysRevB.96.041122}. Interestingly, the studies found that within the framework of non-equilibrium dynamics, as thermalization occurs, the system can evolve from an asymmetric initial state to a symmetric equilibrium state.
To better understand how such symmetry restoration emerges during thermalization, Filiberto Ares, Sara Murciano and Pasquale Calabrese first introduced a quantified indicator for asymmetry named \emph{entanglement asymmetry} (EA) \cite{Ares2023}. This quantity can effectively measure how far an arbitrary reduced density matrix of the subsystem is from being symmetric in an isolated quantum system, allowing us to gain a deeper understanding of the mechanism for restoring symmetry in the subsystem.

\begin{figure}[!th]
    \centering    \includegraphics[width=0.5\linewidth]{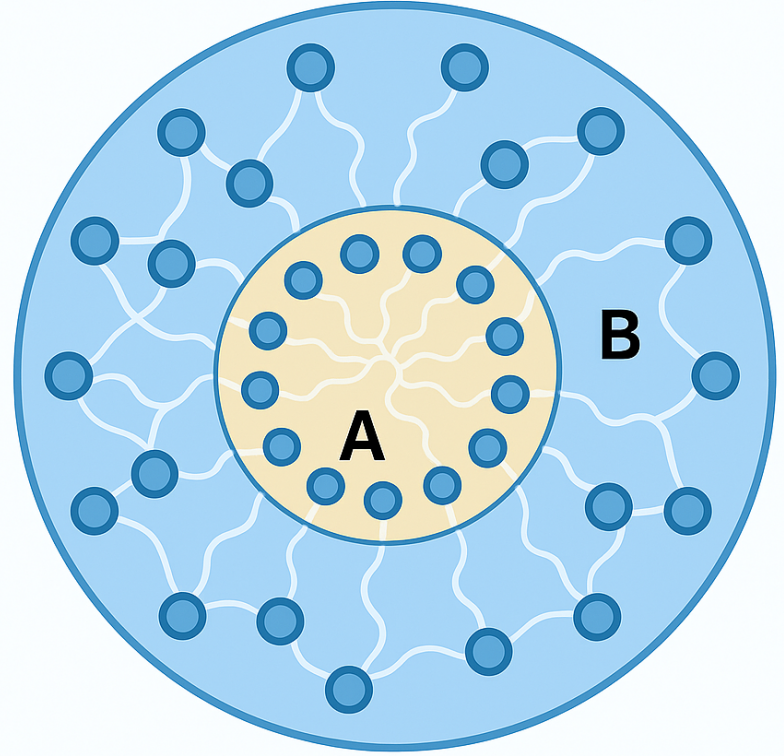}
    \caption{\RaggedRight  Schematic illustration of bipartition of the complex SYK model into subsystem $A$ and its complement $B$.}
    \label{schematic_diagram}
\end{figure}

For a closed Hermitian system, any pure state obeys unitary evolution and remains pure, and the von Neumann entropy of the whole system is thereby zero at any time. However, upon a bipartition into subsystems $A$ and $B$, as the schematic shown in fig.~\ref{schematic_diagram}, even if one considers the initial state, which is composed of the direct product of the pure states of two systems, the information spreading and scrambling between subsystems will cause the reduced state of the subsystem to become mixed. As a result, the entanglement entropy starts from zero and increases with the time evolution. Furthermore, the eigenstate thermalization hypothesis (ETH) \cite{D'Alessio03052016,Deutsch_2018,Gogolin_2016,RevModPhys.76.1267} indicates that in a sufficiently complex and well-coupled system, the local observables of subsystems will tend towards the expectation of a certain thermal state distribution with the time evolution and its density matrix tends to a thermal density matrix after reaching equilibrium. As explained in \cite{Liushuo2024}, the density matrix of a thermal state is naturally block diagonal under the eigenbasis of the conserved charge $\hat{Q}$. Hence, an initially U(1) asymmetric state, whose $\rho_A(0)$ carries coherences across distinct charge sector contributions, undergoes a form of subsystem-level symmetry restoration as the thermalization proceeds. During this process, the asymmetry indicator EA decays to zero.

In this work, we are particularly interested in how symmetry will be restored as the system becomes thermalized in a strongly correlated disordered quantum many-body system. The complex Sachdev-Ye-Kitaev (cSYK) model, a fast scrambling model \cite{Yasuhiro_Sekino_2008, Maldacena_2016,PRXQuantum.5.010201,PhysRevLett.126.030602,PhysRevD.103.106023} with global U(1) symmetry which is considered to be dual to the 2D Einstein-Maxwell dilaton gravity theory, provides us an ideal testbed \cite{PhysRevB.94.035135,PhysRevB.95.155131,Gu2020,Bulycheva2017,PhysRevB.107.075132,PhysRevD.108.086014,ge2025,Ge_2020,cao2024}. Here, we use the exact diagonalization (ED) method to show a detailed demonstration of the symmetry restoration mechanism of the cSYK model. 

Meanwhile, during the process of the restoration of symmetry in quantum systems, a counterintuitive effect can emerge, namely the quantum Mpemba effect (QME) \cite{PhysRevLett.134.220403,PhysRevLett.133.010402,PhysRevLett.133.010401,Zhang2025,PhysRevLett.131.080402,PhysRevA.110.022213,summer2025}. It originates from the classic Mpemba effect, which states that a hot object can cool down faster than a cold object in the same environment \cite{Mpemba_1969,Burridge2016}. The quantum version of Mpemba effect describes the phenomenon that a quantum initial state that is further away from the target state is able to approach the thermal equilibrium or restore symmetry faster. By considering the cSYK model at finite $N$ system, we specifically investigated how QME emerges in the cSYK model. Combined with exact diagonalization numerical methods, the QME during the symmetry restoration process is demonstrated, that certain initial states that are more deviated from U(1) symmetry exhibit an accelerated return in the thermalization process, where the ``farthest ones arrive first'' phenomenon occurs due to the weaker cross-charge sectors coupling in the slowest symmetry restoration mode.

\section{complex SYK model and global U(1) symmetry}\label{sec:model}
\subsection{The model}
In this work, we focus mainly on the phenomenon of symmetry restoration of the complex SYK model $q=4$. The Hamiltonian is
\begin{align}
    H=\sum_{i,j,k,l=1}^N J_{ijkl} c_i^\dagger c_j^\dagger c_k c_l-\mu\sum_{i=1}^N c_i^\dagger c_i.
\end{align}
where $\mu$ is the chemical potential, to simplify the problem, in this work we will focus on the case where the chemical potential is zero. The neutral case is sufficient for studying the behavior of the system's symmetry restoration. This spinless model is $0+1$ dimensional. All the fermion sites are all-to-all randomly coupled by the random variables $J_{ijkl}$ which satisfy the Gaussian distribution with
\begin{align}
\overline{J_{ijkl}}=0, \quad\overline{J_{ijkl}^2}=J^2/8N^3,\nonumber\\
J_{ijkl}=-J_{jikl}=-J_{ijlk}=J_{klij}^*.
\end{align}
The symmetry constraints for $J_{ijkl}$ here are to ensure the Hermitian property of the model. Furthermore, the Hamiltonian is invariant under the transformation $c_i\rightarrow e^{-i\phi} c_i$, indicating that the system has a global U(1) symmetry. The corresponding conserved charge reads
\begin{align}\label{conserved_charge}
    \hat{Q}=\sum_{i=1}^N c_i^\dagger c_i, \quad [H,\hat{Q}]=0.
\end{align}

With the global U(1) symmetry, the Fock space of the Hamiltonian can be divided into a series of direct sum of the subspaces according to different charge sectors
\begin{align}
\mathcal{H} = \bigoplus_{q=0}^{N} \mathcal{H}_q, \qquad
\dim \mathcal{H}_q = \binom{N}{q},
\end{align}
and in an appropriate basis, it exhibits a block diagonal form.

In addition, cSYK model is a disordered system with random coupling. The observables of concern in physics are usually taken as disorder averages. Let ${J_{ijkl}}$ represent a set of coupling constants that satisfy the given Gaussian distribution and symmetry constraints, and denote $P(J_{ijkl})$ as its probability density. For any observable $O(t;J_{ijkl})$ given by time evolution, the disorder average is defined as
\begin{align}
    \overline{O}(t)&=\mathbb{E}_{\{J\}} [O(t;J_{ijkl})]\nonumber\\&=\int \prod_{ijkl}J_{ijkl}P(J_{ijkl})O(t;J_{ijkl})(t)\nonumber\\
    &\approx \frac{1}{R}\sum_{r=1}^R O^{[r]}(t),
\end{align}
where $O^{[r]}(t)$ is the $r$'th independent numerical realization of $\{J_{ijkl}\}$ from the exact diagonalization.

\subsection{Exact diagonalization}\label{sec:ED}

In order to be able to specifically analyze the nature of the subsystems, we adopt a finite-$N$ method called exact diagonalization (ED).  This strategy keeps the U(1) decomposition $\hat Q=\hat Q_A\otimes\mathbb{I}_B+\mathbb{I}_A\otimes\hat Q_B$ at the matrix level, and allows us to compute real-time unitary evolution, reduced states, and EA without any further approximations such as replica trick or large $N$ approximation. In order to perform ED and numerical analysis of the cSYK Hamiltonian, it is convenient to represent the fermionic degrees of freedom in terms of spin operators. However, the fermionic creation and annihilation operators $c_i,~c_i^\dagger$ satisfy the anticommute relation $\{c_i,c_j^\dagger\}=\delta_{ij}$, 
which are not straightforward to implement the representation with a simple-structured matrix. The Jordan–Wigner (JW) transformation \cite{PhysRevB.94.035135} provides an exact mapping between fermions and spins
\begin{align}
    c_i=\sigma_i^-\prod_{j<i}\sigma^z_j~,\quad
    c_i^\dagger=\sigma_i^+\prod_{j<i}\sigma^z_j~.
\end{align}
Under this mapping, the real symmetrical conserved charge operator reads
\begin{align}\label{conserved_charge_JW}   \hat{Q}=\sum_{i=1}^N\frac{\sigma^z_i+1}{2}.
\end{align}
If the whole system is decomposed into two subsystems: subsystem $A$ at the first $m$ sites and subsystem $B$ at the last $N-m$ sites. The total charge operator can be decomposed into
\begin{align}
    \hat{Q}=\hat{Q}_A\otimes \mathbb{I}_B+\mathbb{I}_A\otimes \hat{Q}_B,
\end{align}
where
\begin{align}    \hat{Q}_A=\sum_{i=1}^m\frac{\sigma^z_i+1}{2},\quad \hat{Q}_B=\sum_{i=1}^{N-m}\frac{\sigma^z_i+1}{2}.
\end{align}

In this work, we utilize the sparsity property of the Hamiltonian after the JW transformation and the Krylov subspace algorithm to save a significant amount of computational resources. Further numerical discussions can be found in appendix \ref{app:numerical}.

\section{Entanglement Asymmetry}\label{sec:EA_intro}
\begin{figure}[t]
    \centering
    \begin{subfigure}[t]{0.49\linewidth}
        \centering        \includegraphics[width=\linewidth]{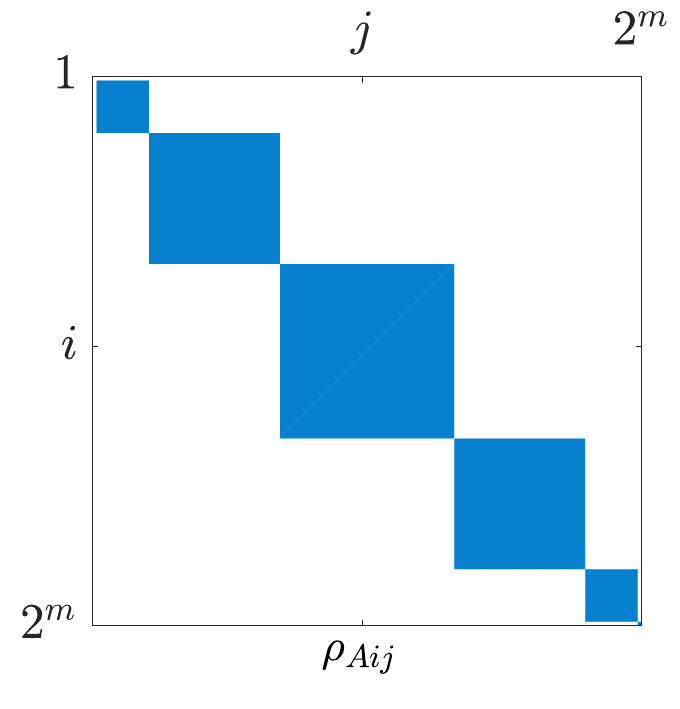}
        \label{fig:rhoA_block}
    \end{subfigure}
    \hfill
    \begin{subfigure}[t]{0.49\linewidth}
        \centering        \includegraphics[width=\linewidth]{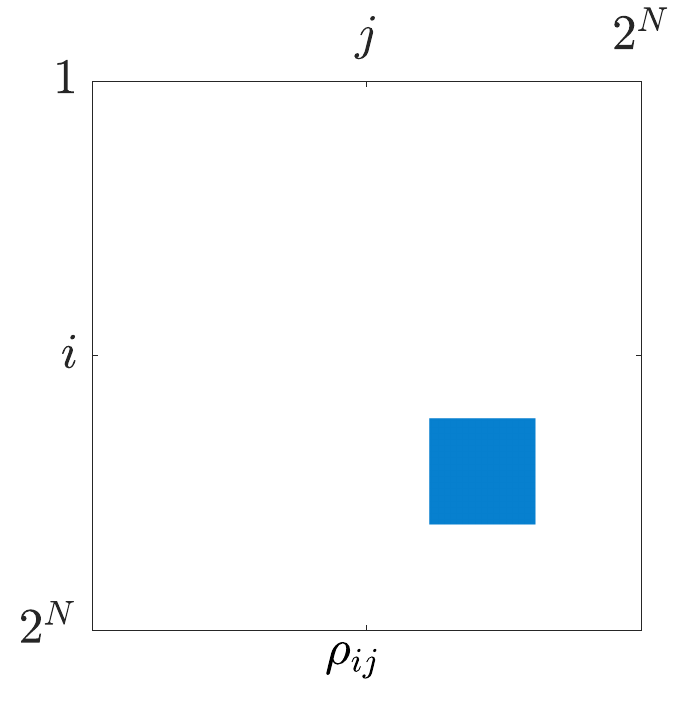}
        \label{fig:rho_block}
    \end{subfigure}    \caption{\RaggedRight Block diagonal structure of the ground state under the single realization of cSYK model, blue region are non-zero matrix elements. (left) Reduced density matrix $\rho_A$ of subsystem $B$ composed by $m=6$ sites. (right) Density matrix $\rho$ of the total system. Here, the number of total sites is $N=12$.}    \label{block_diagonal_structrue}
\end{figure}

To quantify how much a given state breaks the symmetry exactly, Filiberto Ares, Sara Murciano, and Pasquale Calabrese introduce the entanglement asymmetry \cite{Ares2023} as a measurement that is defined as 
\begin{align}\label{EA}
    \Delta S_A=S(\rho_{A,Q})-S(\rho_A),
\end{align}
where $S$ is von Neumann entropy $S(\rho)=-\Tr(\rho \ln \rho)$ and $\rho_{A,Q}=\sum_{q}\Pi_q\rho_A\Pi_q$. Here, $\Pi_q$ are the projectors which project the operator onto the $q$-particles sub-sector. They are the idempotent operators satisfying the relation
\begin{align}
     \Pi_q^\dagger=\Pi_q,~~ \Pi_q\Pi_{q^\prime}=\delta_{qq^\prime}\Pi_q,~~ \sum_q \Pi_q=\mathbb{I},~~\hat{Q}=\sum_q q\Pi_q.
\end{align}
$\rho_{A,Q}$ actually only retains the contribution of the same sector blocks $\Pi_q\rho_{A}\Pi_q$ in $\rho_A$ and sets all crossing sectors $\Pi_q \rho_A\Pi_{q^\prime}=0~~(q\neq q^\prime)$. Therefore, for those states that have no coherent contributions on crossing sectors, we consider such states to be U(1) symmetric, in this case $\rho_{A,Q}=\rho_A$. In \cite{Ares2023}, it has already been indicated that entangled asymmetry can be regarded as the relative entropy
between $\rho_A$ and $\rho_{A,Q}$. We provide a quick proof to show this in appendix \ref{EA=RE}.

One important feature is that if a state is the eigenstate of the charge operator $\hat{Q}$, then the reduced density matrix of the subsystem commutes with the charge operator of the subsystem $[\rho_A,\hat{Q}_A]=0$. 
Notably, the reduced density matrix $\rho_A$ will automatically possess a block diagonal structure under the eigenbasis of the charge operator $\hat{Q}_A$. To see this in the cSYK model, we consider the ground state of the total system derived from the single numerical realization. As discussed in \ref{sec:ED}, we identify the first $m$ sites in the total system as subsystem $B$, and consider the $N-m$ sites left as subsystem $B$. After partial tracing the freedom of subsystem $B$, one obtains the reduced density matrix of subsystem $A$
\begin{align}
    \rho_A=\Tr_B(\rho).
\end{align}
In fig.~\ref{block_diagonal_structrue}, we display the non-zero elements $\rho_{ij}$ and $\rho_{Aij}$ of the cSYK model, which show a clear block diagonal distribution. Furthermore, the ground state of the Hamiltonian is also the eigenstate of the particle number operator $\hat{Q}$. Under the eigenstate representation of $\hat{Q}$, the density matrix $\rho$ only occupies particular sectors corresponding to the subspaces with specific charges. Whereas the reduced density matrix is distributed in all sectors.

\subsection{The initial state}
In the previous studies \cite{Ares2023}, the tilted ferromagnetic state can be adopted as the pure state of the total system to study the symmetry recovery behavior of the system
\begin{align}\label{tilted_ferromagnetic_state}
    \ket{\psi(0)}=\ket{\theta; \nearrow \nearrow \cdots}=e^{-i\frac{\theta}{2}\sum_j \sigma_j^y}
\ket{\uparrow \uparrow \cdots}.
\end{align}
Such setup provides a controllable parameter to adjust the degree of asymmetry of the reduced density matrix. Specifically, when $\theta=n\pi, ~\Delta S_A=0$, this state preserves U(1) symmetry, while for the $\theta\neq n\pi$, $\Delta S_A>0$, the symmetry of the state is broken and the maximum of $\Delta S_A$ locates at $\theta=(n+1/2)\pi$.

In the research of cSYK model, one can still adopt such a state. In Fock basis of fermions, the ferromagnetic state $\ket{\uparrow\uparrow\cdots}$ corresponds to the full fermion excited state $\ket{11\cdots}_F=\prod_i c_i\ket{{\rm vacuum}}$. In this work, we will still follow this convention to study the symmetry restoration phenomenon in the cSYK model. As a comparison, we include more discussions on eigenstates as a type of different setup in appendix \ref{eigen_state_EA}, where we explain that the eigenstates of the Hamiltonian are U(1) symmetric and maintain the symmetry. 

For the conserved charge defined in \eqref{conserved_charge}, the state \eqref{tilted_ferromagnetic_state} is its eigenstate only when $\theta=n\pi,~n\in\mathbb{Z}$. It's straightforward by using the conserved charge expression under the JW transformation. For any arbitrary $\theta$, the Pauli matrices $\sigma_j^a$ satisfy
\begin{align}
    e^{-i\theta\sigma_j^a/2}=\cos(\theta/2)-i\sin(\theta/2)\sigma_j^a,
\end{align}
express \eqref{tilted_ferromagnetic_state} in terms of tensor product as
\begin{align}
\ket{\psi(0)}=e^{-i\frac{\theta}{2}\sum_j \sigma_j^y}
\ket{\uparrow \uparrow \cdots}=
\begin{bmatrix}
\cos(\theta/2) \\
\sin(\theta/2)
\end{bmatrix}^{\otimes N},
\end{align}
apply the conserved charge operator \eqref{conserved_charge_JW} on the state
\begin{align}
    &\hat{Q}\ket{\psi(0)}=    \left(\sum_{i=1}^N\frac{\sigma^z_i+1}{2}\right)\begin{bmatrix}
\cos(\theta/2)  \\
\sin(\theta/2) 
\end{bmatrix}^{\otimes N}\nonumber\\=&\sum_i
    \begin{bmatrix}
\cos(\theta/2)  \\
\sin(\theta/2) 
\end{bmatrix}^{\otimes(i-1)}\otimes \begin{bmatrix}
\cos(\theta/2)  \\
0 
\end{bmatrix}\otimes\begin{bmatrix}
\cos(\theta/2)  \\
\sin(\theta/2) 
\end{bmatrix}^{\otimes(N-i)}.
\end{align}
When $\theta = n\pi,~n\in\mathbb{Z}$ is $\ket{\psi(0)}$ the eigenstate of $\hat{Q}$, in other words, U(1) symmetrical. Besides,
\begin{align}
    &\hat{Q}\ket{\psi(0)}=0,~&{\rm for}~\theta=(2m+1)\pi,~m\in\mathbb{Z}\nonumber\\    &\hat{Q}\ket{\psi(0)}=N\ket{\psi(0)},~&{\rm for}~\theta=2m\pi,~m\in\mathbb{Z}.
\end{align}

\section{Symmetry restoration and thermalization}\label{sec:S_and_EA}
\subsection{Quench on states}
\begin{figure}[!th]
    \centering    \includegraphics[width=\linewidth]{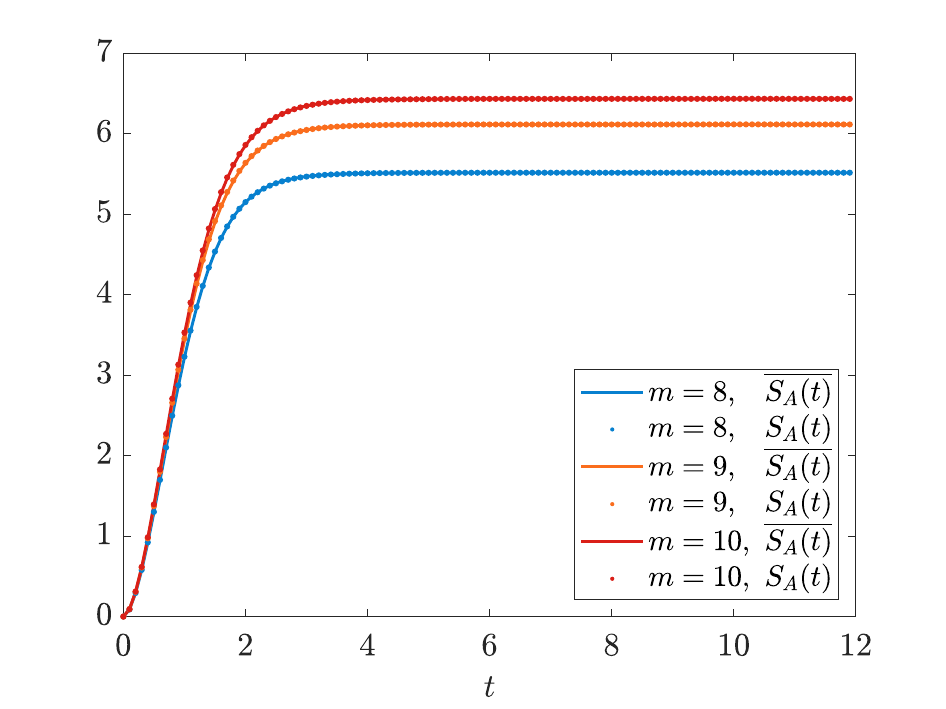}
    \caption{\RaggedRight  The 50 times realizations average entanglement entropy (solid line) and its non-averaged single realization (dotted) as a function of time with various subsystem lengths $m$, where $\theta=\pi/2,~N=20$.}
    \label{fig:S_t}
\end{figure}

In the framework of Hermitian quantum mechanics, the initial state obeys a unitary evolution
\begin{align}
    \ket{\psi(t)}=e^{-iHt}\ket{\psi(0)}.
\end{align}
For a closed system, the density matrix obeys the Liouville–von Neumann equation
\begin{align}
    i\frac{d\rho(t)}{dt} = [H, \rho(t)],\quad \rho(t) = e^{-iHt} \, \rho(0) \, e^{iHt}.
\end{align}
However, the evolution of reduced density matrices $\rho_{A/B}$ is not unitary. In fig.~\ref{fig:S_t}, we show the entanglement entropy $S(\rho_A(t))$ as a function of time with various subsystem lengths $m$. The initial state is chosen to be the maximum tilted ferromagnetic state $\ket{\theta=\pi/2; \nearrow \nearrow \cdots}$ of \eqref{tilted_ferromagnetic_state}. This state is a direct product of two pure states between subsystems at $t=0$. The vanishing entanglement entropy at the origin in this figure well confirms this. Over time, random interactions from the cSYK model drive the rapid diffusion of information in the system, which exhibits fast scrambling. This short-term upward range precisely reflects the dynamic characteristics of information scrambling and entanglement generation. In the long-time limit,  when $S_A(t)$ reaches a stable saturation value, it indicates that the subsystem tends toward an equilibrium. The local observations of the subsystem can be described by a certain thermal density matrix, which is exactly the manifestation of the ETH at the subsystem level. Notice that as $m$ increases, the final saturation height of entropy also rises accordingly, satisfying the volume law. We also demonstrate that there is basically no difference between a single realization $S_A(t)$ (the dotted line) and an average $\overline{S_A(t)}$ of 50 times realizations. It indicates that at the size of $N=20$, the system has already demonstrated a strong self-averaging property.

\begin{figure}[!th]
    \centering    \includegraphics[width=\linewidth]{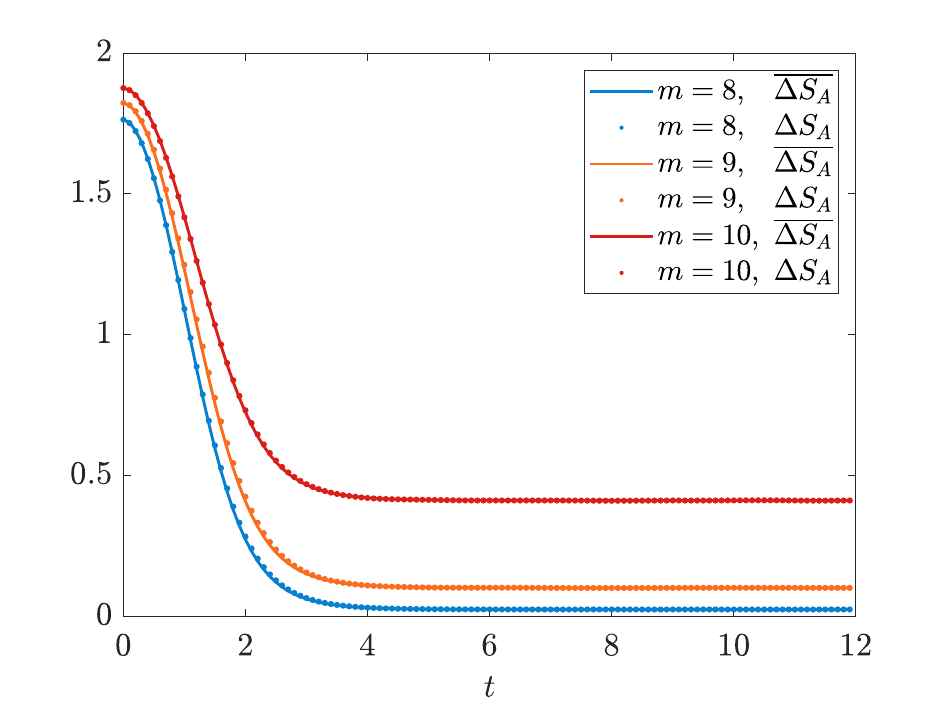}
    \caption{\RaggedRight  The 50 times realizations average entanglement asymmetry (solid line) and its non-averaged single realization (dotted) as a function of time $\Delta S_A(t)$ with various subsystem lengths $m$, where $\theta=\pi/2,~N=20$.}
    \label{fig:EA_t}
\end{figure}

We further calculate EA at multiple lengths $(m\leq N/2)$, as shown in fig.~\ref{fig:EA_t}, $\Delta S_A(t)$ decreases monotonically from a larger initial value and quickly reaches a steady plateau, which is similar to the entanglement entropy. This behavior indicates the coherence crossing sectors of $Q_{A}$ are quickly erased by the environment $B$. It can be clearly observed that as $m$ decreases, or equivalently, the ratio of $m/N$ decreases, the platform height of $\Delta S_A(t)$ significantly decreases in the late time. This means that the smaller $m$ is, the larger the heat bath $B$ with $N-m$ length is, and the U(1) symmetry of subsystem $A$ will be restored more completely. When $m/N=8/20$, $\Delta S_A(t)$ basically vanishes in the late time, that is, the subsystem recovers the symmetry. As for EA $(m=9,10)$ with certain residuals at late time. Its plateau height $\Delta S_A(\infty)$ indicates that the subsystem has only reached equilibrium but not complete thermalization. Therefore, EA can be regarded as an important indicator for measuring the degree of subsystem thermalization with U(1) symmetry.

\subsection{The lower bound of entanglement asymmetry}
\begin{figure}[!th]
    \centering    \includegraphics[width=\linewidth]{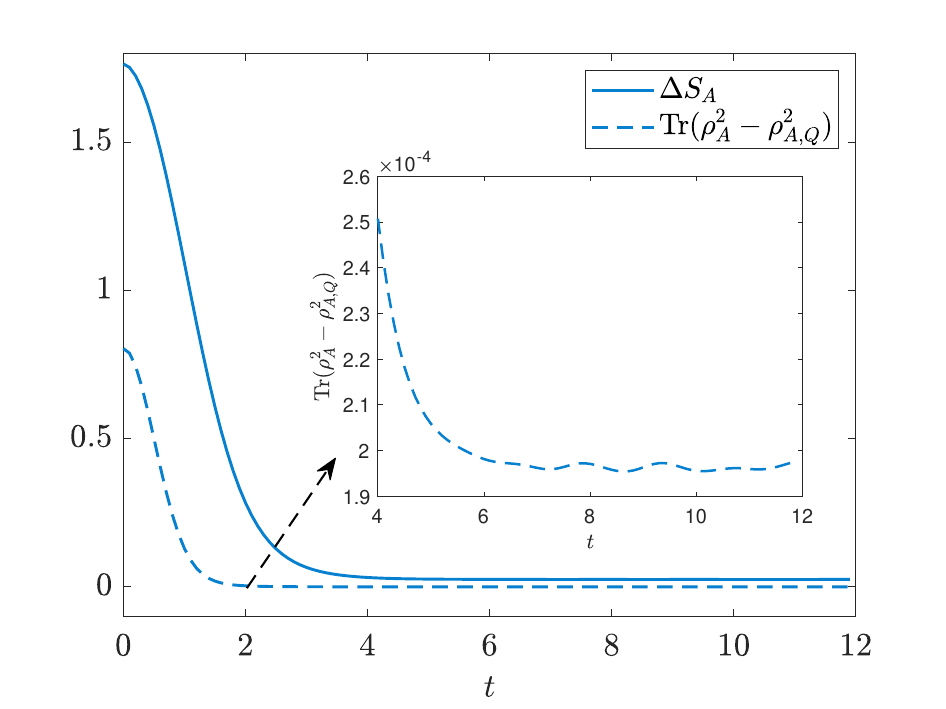}
    \caption{\RaggedRight  The Pinsker-type bound (dotted) of entanglement asymmetry, i.e. purity difference $\Tr(\rho_A^2-\rho_{A,Q}^2)$ and entanglement asymmetry (solid line) in a single realization at $m=8,~\theta=\pi/2,~N=20$.}\label{fig:pinsker_lower_bound}
\end{figure}

The late time height $\lim_{t\rightarrow\infty}\Delta S_A(t)$ of EA can be further analyzed through Pinsker's inequality \cite{King_1998,Carlen_2014,hirota2020}
\begin{align}
S(\rho||\sigma):=\Tr(\rho(\ln\rho-\ln\sigma))\geq\frac{1}{2\ln2}||\rho-\sigma||_1^{~2},
\end{align}
where $||\rho-\sigma||_1$ is the trace norm of $\rho-\sigma$. This inequality determines a lower bound of relative entropy. For EA, we have
\begin{align}\label{lower_bound}
    \Delta S_A\geq\frac{1}{2\ln2}||\rho_A-\rho_{A,Q}||_2^2=\Tr(\rho_A^2-\rho_{A,Q}^2),
\end{align}
where we use the fact of Matrix norm $||X||_1\geq||X||_2$. The square of the matrix norm $||X||_2$ is essentially the sum of the absolute squares of all matrix elements $\sum_{ij}|X_{ij}|^2$. $\Tr(\rho_A^2)-\Tr(\rho_{A,Q}^2)$ stands for the purity difference between $\rho_A$ and $\rho_{A,Q}$. This inequality indicates that EA will not be lower than the purity difference between the two density matrices. If the subsystems reach the maximum entanglement, we have $\Tr(\rho_A^2)=\Tr(\rho_{A,Q}^2)=2^{-m}$ and this lower bound is zero. However, in the long-time limit, if the subsystems cannot reach the maximum mixed state $\rho_A=\mathbb{I}/2^{m}$ through the unitary evolution $U(t)$ of the whole system, this vanishing lower bound is no longer guaranteed. This Pinsker-type lower bound strictly limits the plateau that EA can reach in the late time mathematically. As shown in fig.~\ref{fig:pinsker_lower_bound}, by comparing the same configuration curve, it can be seen that at any time, EA will never be less than the purity difference between $\rho_{A,Q}$ and $\rho_A$, that is, the lower bound defined in \eqref{lower_bound} is always satisfied. Here, we show the non-disorder averaged case since once the single realization is true, the disorder averaged EA automatically holds.

\section{Quantum Mpemba effect}\label{sec:QME}

In the previous section, we show that EA of the subsystem monotonically decreases under unitary dynamics, signaling the restoration of the U(1) symmetry at late times. In this section, we investigate how this restoration depends on the asymmetry of the initial state. We use as a family of quenches the tilted ferromagnetic states parametrized by an angle $\theta$, which continuously tunes the amount of U(1) asymmetry in the initial state. Motivated by the observation that smaller subsystems achieve lower late-time plateaus, we fix the subsystem size to $m=8$, for which symmetry restoration is pronounced, and study dynamics in a system with $N=20$ modes.

\begin{figure}[!th]
    \centering    \includegraphics[width=\linewidth]{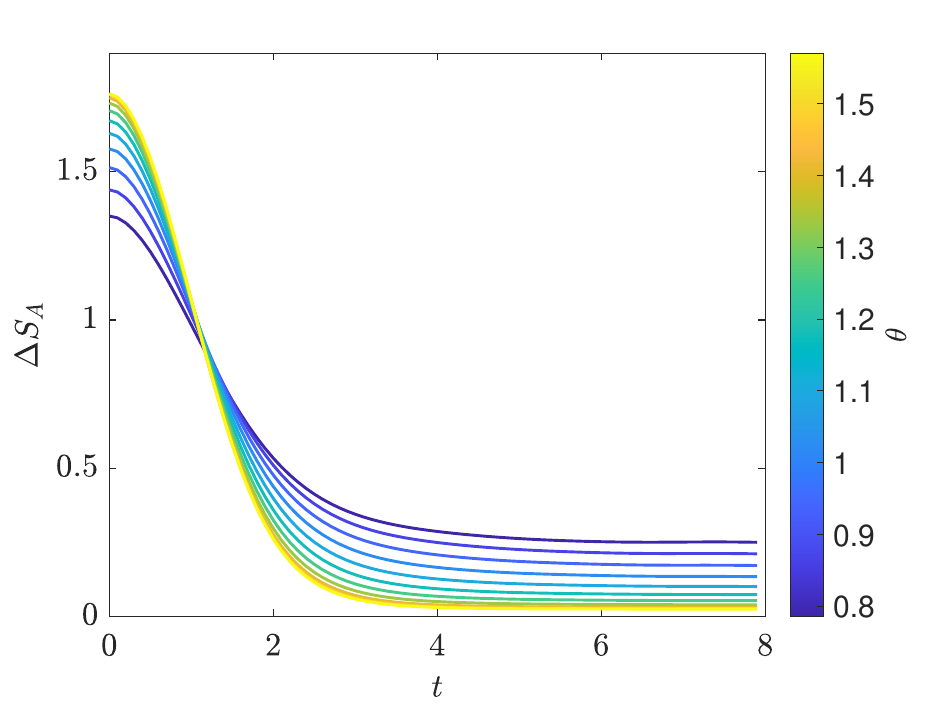}
    \caption{\RaggedRight The averaged entanglement asymmetry of different $\theta\in [\pi/4,\pi/2]$, where $m=8$, $N=20$, derived from 20 times realizations.}
    \label{fig:S_t_theta}
\end{figure}

Fig.~\ref{fig:S_t_theta} displays $\Delta S_A(t;\theta)$ for several values of $\theta\in[\pi/4,\pi/2]$. As $\theta$ increases, the initial asymmetry $\Delta S_A(0;\theta)$ grows, i.e., states are prepared progressively farther from the U(1) symmetric state. Counterintuitively, for more asymmetric initial states, they decrease more rapidly in the early stage (within the range of $t\lesssim2$) and are able to reach a lower plateau at a slightly later time. This is precisely the manifestation of the quantum Mpemba effect in the cSYK model, that is, for an initial state that is more asymmetric, it can restore symmetry at a faster rate. 

Comparing with the Heisenberg XXZ spin chain considered in \cite{Ares2023} where the number of sites is set to be $N = 10$, our numerics of the cSYK model with more sites show a more regularized behavior. That is to say, in both the early and late time stages, the averaged EA of the self-averaging cSYK model can exhibit a relatively steady and monotonically decreasing behavior, and the plateau reached in the late time is also stable without obvious fluctuations. We attribute this behavior to the random all-to-all interactions within the cSYK model. In such a disordered system, the subsystem $A$ is placed in a chaotic thermal bath. The coherence across the $Q_A$ sectors in the reduced density matrix $\Pi_q\rho_A\Pi_{q^\prime}$ is strongly and rapidly shuffled within $O(1)$ time. Therefore, compared to a relatively integrable model, QME in cSYK can exhibit more regular symmetry restoration, and a more stable late-time EA even for a relatively finite sites system.

A useful physical picture is that the tilt parameter $\theta$ controls the initial charge-sector distribution in $A$. Larger $\theta$ prepares broader superpositions across the $Q_A$ sectors, thereby enhancing dephasing between sectors once $A$ entangles with its complement $B$. In a chaotic model such as cSYK, this leads to a faster suppression of the off-block coherences of $\rho_A(t)$ in the $Q_A$ basis and hence a more rapid drop of $\Delta S_A(t;\theta)$. Finite-size effects manifest as a nonzero late-time plateau, whose value decreases for smaller $m$.

\section{Discussion and conclusion}

In this work, we have investigated symmetry restoration and thermalization in cSYK model with a global U(1) symmetry in the framework of finite $N$ exact-diagonalization. Central to our analysis is the entanglement asymmetry, defined as the relative entropy between the reduced density matrix $\rho_A$ and its U(1)–dephased counterpart $\rho_{A,Q}$. Starting from a family of quenched ferromagnetic product states that controllably break U(1), we tracke both the entanglement entropy and EA during unitary evolution. The dynamics display two complementary regimes. First, the system shows a rapidly scrambling information, $S_A(t)$ exhibits a “rise-and-plateau” growth that approaches a volume law, consistent with subsystem ETH and fast scrambling. Second, U(1) symmetry is progressively restored at the subsystem level: EA decays monotonically and saturates at a small late-time value that depends on the subsystem sites $m$. Smaller subsystems (equivalently, larger baths) achieve more complete restoration, with the residual EA essentially vanishing when $m/N$ is sufficiently small. A Pinsker-type lower bound links EA to a purity difference between $\rho_A$ and $\rho_{A,Q}$, clarifying that residual sector coherences—and thus incomplete restoration—are quantitatively suppressed at late times. Taken together, these results portray a consistent picture that entanglement growth thermalizes local observables, while charge-sector dephasing eliminates symmetry-breaking coherences, reconciling fast scrambling with emergent symmetry at long times. The observed entropy curve and its approach to equilibrium echo the eternal black hole Page-curve phenomenology anticipated in holographic settings.

We further uncover a quantum Mpemba effect in symmetry restoration which states prepared farther from the symmetric manifold relax faster in early and intermediate times and reach a lower EA. A natural interpretation is that stronger initial sector superposition enhances dephasing across U(1) charge blocks, accelerating coherence loss.

Methodologically, our ED numerical framework and the EA diagnostic constitute a portable toolkit for symmetry-sensitive non-equilibrium studies.  Looking ahead, extending to other continuous or discrete symmetries, exploring open-system settings, scaling to large $N$, and relating EA dynamics to out-of-time-order correlators may sharpen connections to holography and experiments in programmable quantum simulators.

\acknowledgments
We would like to thank Ze-Hua Tian, Yiheng Lin, and Yu-Qi Lei for valuable discussions. This work is supported by NSFC, China (Grant No. 12275166 and No. 12311540141).

\bibliographystyle{bibstyle}
\bibliography{refs}

\appendix
\section{The sparse property of the Hamiltonian}\label{app:numerical}
The JW transformation converts the Hamiltonian into the form of a spin chain. Its matrix representation is highly sparse. 
\begin{table}[htbp]
  \centering
  \caption{The relationship between the sparsity of the matrix in the cSYK model and $N$}
  \label{tab:csyk_sparsity}
  \begin{tabular}{c c}
    \hline
    $N$ & \quad\# non-zero elements/$4^N$ \\
    \hline
    12 & 5.36\% \\
    13 & 3.76\% \\
    14 & 2.57\% \\
    15 & 1.73\% \\
    16 & 1.13\% \\
    17 & 0.73\% \\
    18 & 0.47\% \\
    \hline
  \end{tabular}
\end{table}

The sparsity of the Hamiltonian with different sites number $N$ is shown in table \ref{tab:csyk_sparsity}. At $N=12$, the non-zero matrix elements account for only 5.36\%. When $N = 18$, this proportion further decreases to 0.47\%. The sparsity significantly decreases as $N$ increases. This characteristic can significantly reduce memory costs of computing by using sparse matrix storage, which only records the positions and values of the non-zero elements in the matrix. Furthermore, since we are interested in the unitary evolution of a pure state, which essentially is an e-exponential operator acting on a state $\hat{U}(t)\ket{\psi(0)}$, this allows us to employ the Krylov subspace method \cite{Nandy_2025,Saad1992,Liesen2012} to significantly accelerate the calculation of the time evolution. This algorithm demonstrates greater superiority for matrices that are sparser and enables us to obtain $\ket{\psi(t)}$ from the initial state $\ket{\psi(0)}$ with much fewer memory usages, without generating the specific time evolution operator that requires $4^N\times16$ Bytes storage.

\section{Relative entropy form of entanglement asymmetry}\label{EA=RE}
In this appendix, we prove how entanglement asymmetry defined in \eqref{EA} is essentially the relative entropy $S(\rho_A||\rho_{A,Q})$ between $\rho_A$ and $\rho_{A,Q}$. Notice that
\begin{align}
    \Tr(\rho_{A,Q}\ln \rho_{A,Q})&=\sum_q\Tr(\Pi_q\rho_A\Pi_q\ln \rho_{A,Q})\nonumber\\    &=\sum_q\Tr\left(\rho_A(\Pi_q\ln \rho_{A,Q}\Pi_q)\right)\nonumber\\
    &=\Tr\left(\rho_A\ln \rho_{A,Q}\right),
\end{align}
here we use the fact that $\sum_q\Pi_q\ln \rho_{A,Q}\Pi_q=\ln \rho_{A,Q}$. By the definition of entanglement entropy, we have
\begin{align}
    \Delta S_A&=-\Tr(\rho_{A,Q}\ln \rho_{A,Q})-\Tr(\rho_{A}\ln \rho_{A})\nonumber\\
    &=\Tr\left(\rho_A(\ln \rho_{A}-\ln\rho_{A,Q})\right)\nonumber\\    &=S(\rho_A||\rho_{A,Q}),
\end{align}
where $S(\rho_A||\rho_{A,Q})$ represents the relative entropy between $\rho_A$ and $\rho_{A,Q}$. It naturally inherits all properties of relative entropy, such as its non-negativity, asymmetry $S(\rho_A||\rho_{A,Q})\neq S(\rho_{A,Q}||\rho_{A})$, unitary invariance $S(\rho_A||\rho_{A,Q})=S(U \rho_AU^\dagger||U\rho_{A,Q}U^\dagger)$, Pinsker inequality $S(\rho||\sigma)\geq ||\rho-\sigma||_1^2/(2\ln2)$, etc. 

\section{Entanglement Asymmetry of the eigenstates of Hamiltonian}\label{eigen_state_EA}
In this appendix, when setting the eigenstates as the initial setup, we show that for the eigenstates of the Hamiltonian, any subsystem maintains a U(1) symmetric state at all times.

For the set of eigenstates of the Hamiltonian $\{\ket{E_n} \}$, the eigenstates obey the unitary time evolution $U(t)=e^{-iHt}$, and the density matrix is invariant under time evolution
\begin{align}
    \rho(t)=e^{-iHt}\ket{E_n}\bra{E_n}e^{iHt}=\ket{E_n}\bra{E_n}=\rho(0).
\end{align}
Therefore, the reduced density matrix of any subsystem is also time-independent. Next, we show that the eigenstates of the Hamiltonian are simultaneously the eigenstates of the charge operator. The non-degenerate state case is straightforward. From $[H,Q]$ we have
\begin{align}    HQ\ket{E_n}=QH\ket{E_n}=E_nQ\ket{E_n}.
\end{align}
This means that the state $Q\ket{E_n}$ is also the eigenstate of the Hamiltonian, $Q\ket{E_n}$ lies only in the subspace with energy level of $E_n$. The non-degeneracy implies the dimension of the eigenspace is one, $Q\ket{E_n}$ can only be a linear scaling of $\ket{E_n}$, thereby $Q\ket{E_n}=q\ket{E_n}$, i.e. $\ket{E_n}$ is the eigenstate of $Q$ with eigenvalue $q$. As for eigenstates with a certain degree of degeneracy, in mathematics, one can always choose a basis that diagonalizes both $H$ and $Q$ simultaneously. For the basis vectors like this, they are still the eigenstates of $Q$.

It has been already shown that $[\rho_A,Q_A]=0$ for the eigenstate $\ket{\Psi}$ of $Q$ in \cite{Ares2023}. Here we briefly prove this. Let
\begin{align}
    U(\alpha)=e^{i\alpha Q}=e^{i\alpha Q_A}\otimes e^{i\alpha Q_B},\quad \alpha \in\mathbb{R}.
\end{align}
For the eigenstate $Q\ket{\Psi}=q\ket{\Psi}$ we have $U(\alpha)\ket{\Psi}=e^{i\alpha q}\ket{\Psi}$, hence
\begin{align}
    U(\alpha)\ket{\Psi}\bra{\Psi}U(-\alpha)=\ket{\Psi}\bra{\Psi}.
\end{align}
Partial tracing of subsystem $B$ we derive
\begin{align}
\rho_A&=\Tr_B(\ket{\Psi}\bra{\Psi})=\Tr_B\left(U(\alpha)\ket{\Psi}\bra{\Psi}U^\dagger(\alpha)\right)\nonumber\\
&=e^{i\alpha Q_A}\Tr_B]\left(\ket{\Psi}\bra{\Psi}\right)e^{-i\alpha Q_A}\nonumber\\
&=e^{i\alpha Q_A}\rho_A e^{-i\alpha Q_A}.
\end{align}
Taking the derivative of $\alpha$ at $\alpha=0$
\begin{align}
    0=\frac{d}{d\alpha}\left(e^{i\alpha Q_A}\rho_A e^{-i\alpha Q_A}-\rho_A\right)\bigg|_{\alpha=0}=i[Q_A,\rho_A].
\end{align}
$[Q_A,\rho_A]=0$ implies they can be simultaneously diagonalized, which has a block diagonal structure under the spectral projection $\{\Pi_q\}$ of $Q_A$, meanwhile $\Delta S_A=0$. For the time-independent density matrix $\rho$ and $\rho_A$ of the non-degenerate eigenstates of the Hamiltonian, the entanglement asymmetry remains zero. Indicating that these eigenstates maintain U(1) symmetry after the quench.

\end{document}